\documentclass{JHEP3}

\usepackage{amsmath}
\usepackage{amsfonts}

\def\be{\begin{equation}}
\def\ee{\end{equation}}
\def\ba{\begin{eqnarray}}
\def\ea{\end{eqnarray}}
\def\bs{\begin{subequations}}
\def\es{\end{subequations}}

\title{de Sitter thermodynamics and the braneworld}
\author{Gianluca Calcagni\\
Department of Physics, Gunma National College of Technology, Gunma 371-8530, Japan\\
Astronomy Centre, University of Sussex, Brighton BN1 9QH, U.K.\\
E-mail: \email{calcagni@nat.gunma-ct.ac.jp}}

\abstract{The de Sitter thermodynamics of cosmological models with a modified Friedmann equation is considered, with particular reference to high-energy Randall-Sundrum and Gauss-Bonnet braneworlds. The Friedmann equation can be regarded as the first law of thermodynamics of an effective gravitational theory in quasi de Sitter spacetime. The associated entropy provides some selection rules for the range of the parameters of the models, and is proposed for describing tunneling processes in the class of high-energy gravities under consideration.}

\preprint{JHEP09(2005)060 \hspace{2cm} hep-th/0507125}

\begin{document}


\section{Introduction}

Due to the progress in the understanding of string theory, in the last few years much attention has been devoted to braneworld cosmological models in a higher-dimensional spacetime. The braneworld scenario was soon recognized as a gravity system coupled to a conformal field theory (CFT) \cite{kra99,gub01}. In fact, it has an immediate and surprising holographic description through the Cardy-Verlinde formula, relating the entropy of a CFT to its central charge and hamiltonian \cite{car86,ver00}. A brane with matter in an anti de Sitter (AdS) background ca be viewed as an empty brane in a Schwarzschild- anti de Sitter (Sc-AdS) bulk, onto which the bulk black hole thermodynamics induces a field theory which is conformal if the brane is at or close to the boundary of the spacetime, and nonconformal otherwise. The Friedmann equation describing the effective cosmological evolution of the brane coincides with the Cardy-Verlinde formula at the instant when the brane, during its expansion from the inner region of the black hole, crosses the black hole horizon. This braneworld holography was studied in \cite{cvrs,GrP2,MyK} considering Einstein gravity (Randall-Sundrum case), while the case of black holes and holography in Gauss-Bonnet or Lovelock gravity was analyzed, e.g., in \cite{gbbh,NO02,GrP} (see \cite{MyS} for black hole thermodynamics in this class of models).

The basic assumption implicit in the AdS/CFT description is that the AdS radius $a(t)$ where the brane is located (respectively, the size of the manifold the CFT lives on) is larger than the curvature radius $\ell$ of AdS (the length scale of the thermal CFT), $\ell/a\ll 1$, so that the CFT is at high temperature. Since this stage of the brane evolution corresponds, from a four-dimensional point of view, to a low-energy regime, the Friedmann equation resulting from the boundary Cardy-Verlinde formula is of standard type, 
\be\label{gr}
H^2 = \beta_1^2\rho,
\ee
where $H\equiv \dot{a}/a$ is the Hubble parameter ($a(t)$ is identified with the scale factor and a dot denotes derivative with respect to synchronous time), $\beta_1^2\equiv\kappa^2_{\rm 4D}/3$, $\kappa^2_{\rm 4D}$ is the 4D gravitational coupling, and $\rho$ is the effective energy density on the brane.

However, at sufficiently high energies the evolutionary path of braneworld cosmologies can differ from that dictated by general relativity. This regime is achieved provided the brane energy density is comparable or larger than the energy required to stabilize the extra dimension(s) against quantum fluctuations, so that the bulk back-reaction modifies eq. (\ref{gr}). In Einstein gravity one can show that there is an exact holographic description of the Randall-Sundrum (RS) full-brane picture in empty AdS in terms of a nonconformal field theory induced on an empty brane by a bulk black hole \cite{GrP2}. If the effect of the black hole mass (and charge, in the Reissner--Nordstr\"{o}m-AdS case) in the $D$-dimensional bulk dominates over the brane tension as regards the effective $(D-1)$-evolution, then the resulting Friedmann equation is that typical of the high-energy, early braneworld; this corresponds to a brane position close to the AdS center, while near the boundary the AdS/CFT correspondence holds and the usual 4D evolution is recovered. However, it was shown that in Gauss-Bonnet (GB) gravity such duality does not exist neither exactly nor approximately in the high-energy limit, while one can rely on the AdS/CFT picture near the boundary. In other words, one cannot recover the GB Friedmann equation from holographic considerations, since the energy density induced on the brane by the bulk black hole evolves in a different way with respect to that of the nonempty GB braneworld. Then the GB cosmological scenario is equivalent to a field theory in Gauss-Bonnet gravity only in the conformal limit, and not throughout the whole bulk range. The reason is that in order to get a holographic correspondence even at high energies the mass of the bulk black hole should be very small (contrary to the RS case) and the Sc-AdS spacetime would tend to AdS trivially \cite{GrP}. Since this limit corresponds to the weak self-gravity regime, $aH\ll 1$, it cannot describe a cosmological brane.

Whether and when some kind of extended holography for gravitational models exists or not outside the AdS/CFT limit is still an unsolved issue that has to address a number of problems. For instance, at low temperature the Sc-AdS black hole is unstable and there might not be a sensible dual description of the boundary brane. Anyway the above studies suggest that there can be nontrivial answers to this question, if the latter is well posed.

It is known that gravity has another simple holographic description in the case of a de Sitter (dS) background. The Friedmann equation can be regarded as the first law of thermodynamics for the cosmological de Sitter horizon in a $D$-dimensional Einstein \cite{jac95,BR,FrK,pad} or Lovelock gravity \cite{CK}. Here we shall show that it is possible to extend this interpretation to the braneworld case, even the GB one, when the effective dynamics does not necessarily satisfy eq. (\ref{gr}) at high energies. 

The evolution of the universe can be described just by a thermodynamical process at its cosmological horizon, which in four dimensions is a two-dimensional surface. The dS entropy in the low-energy or 4D limit is then proportional to the area of the horizon as in the Bekenstein-Hawking law for black holes \cite{BH} (see \cite{SSV} for a comprehensive review). 

One finds that the geometric configuration of the brane universe induces a correction to the area law which depends on the bulk gravity. Although we shall perform the calculations in an effective theory which is connected only at phenomenological level with the original ones (in particular, a domain wall in Einstein or Gauss-Bonnet gravity), the equivalence between the Friedmann evolution and the adiabatic process at the horizon is realized in a way that looks promising for two intertwined reasons. On one hand, a better comprehension of holographic descriptions of backgrounds with boundaries could be crucial for unravelling some fundamental properties of quantum gravity at high energies, including the role of the braneworld at very early times. On the other hand, a robust definition of the entropy of these systems can be directly related to the birth of the universe and its initial conditions in terms of probability amplitudes, as we shall mention in the end of the paper. We hope that these results will trigger further research in both directions.


\section{Setup}

Our starting point is a high-energy inflationary brane universe filled with a perfect fluid responsible for the accelerated expansion. In the high-energy limit the effective cosmological evolution deviates from the linear behaviour of eq. (\ref{gr}). In particular, we can describe the primordial universe, at least in some finite time interval or energy patch, by the Friedmann equation
\be\label{FRW}
H^2=\beta_q^2\rho^q,
\ee
where $q$ is constant and $\beta_q>0$ is a constant factor with energy dimension $[\beta_q]= E^{1-2q}$. This equation encodes several situations, including, among others, the pure 4D radion-stabilized regime ($q=1$), the high-energy limit of the Randall-Sundrum braneworld ($q=2$), and the high-energy limit of the Gauss-Bonnet scenario ($q=2/3$). It is convenient to introduce the parameter
\be
\theta \equiv 2(1-q^{-1}),
\ee
so that eq. (\ref{FRW}) can be recast as
\be\label{FRW2}
H^{2-\theta}=\beta_q^{2-\theta}\rho.
\ee
Deviations from the 4D case $\theta=0$ will characterize exotic scenarios (braneworlds or modified gravities) according to their magnitude and sign. Because of the local conservation of the energy-momentum tensor on the brane, for a perfect fluid with energy density $\rho$ and pressure $p$, the continuity equation is $\dot{\rho}+3H(\rho+p)=0$ when there is no brane-bulk exchange. Here we have neglected curvature and Weyl terms; we will come back to this point in the final section.


\subsection{Effective action on quasi de Sitter background}

It is instructive to regard eq. (\ref{FRW2}) as the approximation of a pure $D$-dimensional quasi de Sitter model with effective action
\be\label{pact}
{\cal S}_{\rm eff}=\int d^Dx \sqrt{-g}\left[\frac{(b_qR)^{1/q}}{2\kappa^2_q}+{\cal L}_\rho(g)\right],
\ee
where $\kappa_q^2$ is the effective gravitational coupling, $b_q$ is a constant, $R$ is the Ricci scalar of the target spacetime, and ${\cal L}_\rho(g)$ is the lagrangian of the perfect fluid. The metric induced on the brane is the Friedmann-Robertson-Walker (FRW) metric
\be\label{metric}
ds^2=-dt^2+\frac{a^2(t)}{1-kr^2}dr^2+\tilde{r}d\Omega^2_{D-2},
\ee
where $\tilde{r}\equiv a(t)r$, $d\Omega^2_{D-2}$ is the line element of a $(D-2)$-sphere with unitary radius, and $k=0,1,-1$ is the curvature in a flat, closed, and open universe, respectively.

Neglecting $O(\epsilon)$ terms, where $\epsilon\equiv -\dot{H}/H^2$ is the first slow-roll parameter, one has $R\approx D(D-1)H^2$, while the Friedmann equation coming from the least action constraint is eq. (\ref{FRW2}), provided
\be\label{kappa}
\kappa_q^2=\frac{D-2+\theta}{2D}[D(D-1)b_q\beta_q^2]^{1/q}.
\ee
In four dimensions, $\kappa_q^2=(2+\theta)(12b_q\beta_q)^{1/q}/8$, which reduces to $\kappa_1^2=3\beta_1^2=\kappa^2_{\rm 4D}=8\pi G$ for $\theta=0$ and $b_1=1$, where $G$ is the 4D Newton constant. Note that the constant (\ref{kappa}) can be identified to the coupling of the full gravity theory only for suitable choices of the coefficient $b_q$; in particular, one can fix $b_q$ so that $\kappa^2_2 = \kappa^2_{\rm RS}=6\beta_2$ ($b_2=2^6/3$) and $\kappa^2_{2/3} = \kappa^2_{\rm GB}=16\alpha\beta_{2/3}^3$ ($b_{2/3}=\alpha^2 2^8/3^3$), $\alpha$ being the Gauss-Bonnet constant. However, this coefficient is introduced by hand only by phenomenological considerations. Moreover, the gravity lagrangian does not contain other curvature invariants than the Ricci scalar, which one would expect in genuinely Lovelock-type theories like Gauss-Bonnet. Third, it gives the patch Friedmann equation only at lowest order in the slow-roll approximation, under the assumption that $O(\epsilon)$ and higher contributions are negligible. For these and other reasons, one should regard eq. (\ref{act}) as an effective model, describing in $D=4$ a cosmological evolution similar to the high-energy evolution on a 3-brane embedded in a higher-dimensional bulk.

At this point a remark is in order. Actually one can extend eq. (\ref{pact}) to a more general case, for instance replacing the $R^{1/q}$ lagrangian with 
\be\label{Lg}
{\cal L}_g=\frac{1}{2\kappa_q^2}\left(a_1R_{\alpha\beta\mu\nu}R^{\alpha\beta\mu\nu}+a_2R_{\mu\nu}R^{\mu\nu}+a_3R^2\right)^{1/(2q)},
\ee
where $a_i$ are arbitrary coefficients. When $a_1\neq 0 \neq a_2$, one can exactly cancel higher-order slow-roll contributions for suitable choices of the coefficients. To do so, we rewrite the lagrangian (\ref{Lg}) as
\be
{\cal L}_g=\frac{(D-1)^{1/(2q)}}{2\kappa_q^2}H^{2/q}\left[c_1(D-4\epsilon)+c_2\epsilon^2\right]^{1/(2q)},
\ee
where $c_1  = 2a_1+(D-1)a_2+D(D-1)a_3$ and $c_2 = 4a_1+Da_2+4(D-1)a_3$. The Friedmann equation corrected up to first-order slow-roll contributions is 
\ba
&&\left[1-\frac{(D-1)\epsilon}{(D-2+\theta)q}\delta_D\right]H^{2-\theta} = \beta_q^{2-\theta}\rho,\\
&&\delta_D \equiv \frac{c_2}{c_1}+\frac{2}{D-1}\left(3-\frac{4}{D}+\theta\right),
\ea
where now $c_1=D(D-1)b_q^2$ in eq. (\ref{kappa}). The condition $\delta_D=0$ fixes the coefficients of the action, which in four dimensions are
\be
\frac{c_2}{c_1}=-\frac{2(2+\theta)}{3};
\ee
note that they are opposite in sign as long as $\theta>-2$, which is the condition for $\kappa_q^2$ to be positive as long as $b_q>0$. In particular, $c_2<0$. In order to avoid confusion, we stress once again that the lagrangian (\ref{Lg}) is motivated only by phenomenology and has nothing to do, for example, with the full Gauss-Bonnet lagrangian with fixed coefficients $a_2=-4a_1=-4a_3$. In the following we shall consider only the simple $R^{1/q}$ action, eq. (\ref{act}), later showing that this does not result in a loss of generality.


\subsection{Conformal transformation}

The effective action (\ref{pact}) can be rewritten as a scalar theory in Einstein gravity. In general, to a polynomial action
\be\label{act}
{\cal S}=\int d^Dx \sqrt{-g}\left[\frac{f(R)}{2\kappa^2}+{\cal L}_\rho(g)\right],
\ee
one can apply a conformal transformation \cite{mae89}
\be\label{contra}
\bar{g}_{\mu\nu} \equiv e^{\frac{2\gamma \varphi}{D-2}}g_{\mu\nu},
\ee
where
\ba
\varphi &\equiv& \gamma^{-1} \ln f'(R),\label{scaf}\\
\gamma &\equiv& \sqrt{\frac{\kappa^2(D-2)}{D-1}},
\ea
and $f'=\partial f/\partial R$. Then the action becomes
\be
{\cal \bar{S}}= \int d^Dx \sqrt{-\bar{g}}\left[\frac{\bar{R}}{2\kappa^2}-\frac12(\bar{\nabla}\varphi)^2-V(\varphi)+e^{-\frac{D}{D-2}\gamma \varphi}{\cal L}_\rho(e^{-\frac{2}{D-2}\gamma \varphi}\bar{g})\right],
\ee
where
\be\label{phipot}
V(\varphi)= \frac{e^{-\frac{D}{D-2}\gamma \varphi}}{2\kappa^2}[Rf'(R)-f(R)].
\ee
The conformal transformation (\ref{contra}) and the scalar field (\ref{scaf}) are well defined when
\be\label{f'>0}
f'(R)>0.
\ee
In the inflationary universe,
\be
R=(D-1)(D-2\epsilon)H^2>0.
\ee
Hence, for the action (\ref{pact}) the condition (\ref{f'>0}) is satisfied when 
\be\label{range}
b_q>0,\qquad 2-D<\theta<2,
\ee
where we have imposed $\kappa_q^2>0$ to discard the $b_q<0$ case. Therefore $q>2/D$. The scalar field potential is then
\be\label{patpot}
V(\varphi) = -\theta\frac{e^{-\frac{D}{D-2}\gamma \varphi}}{4\kappa_q^2}f(R)\propto -\theta e^{-\frac{2(D-2+\theta)}{\theta(D-2)}\gamma \varphi},
\ee
where we have used the expression for the Ricci scalar
\be
R=\left(\frac{b_q^{1/q}}{q}e^{-\gamma\varphi}\right)^{2/\theta}.
\ee
Since the cosmological dynamics fed by the barotropic fluid prescribes a decreasing Hubble parameter, the natural evolution of the system is such that $\dot{\varphi}>0$ for $\theta>0$ (RS braneworld) and $\dot{\varphi}<0$ for $\theta<0$ (GB braneworld). In the first case the potential (\ref{patpot}) is negative definite and the scalar field climbs up from the instability at $\varphi\to -\infty$ to the asymptotically flat region at $\varphi\to +\infty$. In the second case the field rolls down its potential towards the flat region at $\varphi\to -\infty$. Both these situations seem problematic, the first appearing unphysical and the second leading to a naked singularity and to an instability of the system. However, the patch approximation holds by definition only within a certain time interval during the accelerated early-universe expansion, the GB regime being dominant before the RS one.\footnote{The evolution GB$\to$RS$\to$4D will be assumed so that to encompass all the cases of interest, but of course a direct transition from only one high-energy regime (GB or RS) to general relativity is also possible in principle.} Also, the classical picture presented here is likely to be nontrivially broken at very early times and high curvature. Therefore the instability in the $\theta<0$ case can be interpreted as the natural mark of the transition from the GB to the RS phase. Subsequently, in the $\theta>0$ regime the scalar field will actually start from a negative but finite initial value $\varphi_i < 0$ and have sufficient kinetic energy to climb up its potential.

Note also that in a more complete model the energy regimes are smoothly connected between each other, allowing for other terms in the curvature expansion of the gravitational action. These terms considerably change the shape of the conformal potential in a way such that the above considerations may not hold. The simplest example, outside the braneworld picture, is the quadratic lagrangian ${\cal L}_g=R+cR^2$, $c$ being a constant. In the four-dimensional patch approximation the $\theta=-2$ Lagrangian leads to a constant potential (and entropy), while in the other case there is a local minimum or maximum according to the sign of $c$ \cite{JKM1,JKM2}. 

Therefore in some cases eq. (\ref{FRW}) does not provide a consistent cosmological description, unless one considers other terms ($H^2= \alpha_1\rho^q+\alpha_2\rho+\dots$) as is usually done, for instance, in late-time alternative gravity models (see \cite{car1} and related papers). Conversely, one can constrain theories predicting a modified Friedmann evolution by considerations similar to those above, which are tightly related to the dS thermodynamics.

Eq. (\ref{contra}) is just one of the possible conformal transformations of the monomial theory into Einstein gravity coupled to a scalar field. If $f'<0$, the field $\varphi\propto \ln (-f')$ defines another mapping which allows the previously discarded range of parameters, $q<0$ when $b_q>0$ and $0<q<2/D$ when $b_q<0$. The definition $\varphi\propto \ln |f'|$ retains all the features of the model without constraining its parameters. As in ref. \cite{JKM2} we shall impose the conformal transformation eq. (\ref{contra}) as a selector for the parameters of the theory. This choice is justified by the fact that both the RS and GB braneworld models belong to this particular sector.


\section{Thermodynamics}

We now show that the Friedmann equation follows directly from the first law of de Sitter thermodynamics. This result includes the 4D one and the braneworld case when the effective evolution satisfies the monomial behaviour of eq. (\ref{FRW}). We can write the first law of dS thermodynamics as
\be\label{QTS}
-dE=T dS,
\ee
where an infinitesimal adiabatic variation of the total energy $E$ into a causally connected de Sitter coordinate volume is related to a corresponding change in the de Sitter entropy $S$, where $T$ is the Gibbons-Hawking temperature. 

To calculate the entropy of a de Sitter spacetime subject to the modified gravity eq. $(\ref{act})$, we can consider a black hole filling the causal universe so that its event horizon $r_+$ coincides with the Hubble radius, and identify its entropy with that associated to the de Sitter horizon. The black hole entropy formula in higher-derivative gravities was found in \cite{JKM1,JKM2,MiW,IW,JM,CvP} and reads, in the pure Ricci-scalar case,
\be\label{Sf}
S=\frac{2\pi}{\kappa^2}A_{D-2}f'(R),
\ee
where $A_{D-2}$ is the $(D-2)$-dimensional area of a black hole in the gravitational theory given by eq. (\ref{act}). In $D$ dimensions and at the event horizon, $A_{D-2}=(D-1)\Omega_{D-1}r_+^{D-2}$, where $\Omega_{D-1}=\pi^{(D-1)/2}/\Gamma((D+1)/2)$ is the area of the $(D-1)$-dimensional unit ball. In the patch case, evaluated at the de Sitter horizon $r_+=H^{-1}$,
\be\label{S}
S_q=\frac{(2-\theta)2\pi A_{D-2}}{(D-2+\theta)(D-1)\beta_q^{2-\theta}}H^{-\theta},
\ee
which is the standard area entropy law corrected by a factor proportional to $H^{-\theta}$. Equation (\ref{S}) is nothing but the Bekenstein bound of a system of size $R\sim H^{-1}$, which is saturated by the above black hole in an asymptotically flat space. Since $H^{2-\theta}\sim \rho\sim E R^{D-1}$, then $S_{B}\sim ER \sim \rho R^D\sim H^{-(D-2+\theta)}$. This formula gives some insights as regards the viability of patch cosmological models in four dimensions, $D=4$.
\begin{itemize}
\item States with negative entropy may be discarded because they are unphysical, or they represent an instability of the theory and a phase transition, or also because the adopted formalism, in the framework of classical thermodynamics, does not describe them in a correct way \cite{NO02,inst}. In any case, we can select as physically relevant only models with $-2<\theta<2$, that is $q>1/2$. In particular, phantom-like cosmologies with negative $q$, which may arise in late-time gravities dominated by low-curvature terms, are excluded. This is in agreement with the positivity of the effective gravitational coupling, eq. (\ref{kappa}).
\item The marginal case $\theta=-2$ ($q=1/2$) is peculiar for at least two reasons. First, it is the only model displaying a next-to-leading-order tensor degeneracy in the inflationary consistency relations \cite{cal5}. Secondly,  this cosmology gives a scale-invariant spectrum ($n_s=1$) in the case of power-law inflation, $a=t^n$, irrespective of the value of $n$ \cite{KM}. Since the coupling constant (\ref{kappa}) vanishes when $\theta=2-D$ and the entropy diverges, we discard this case as well.\footnote{The 4D and GB scenarios correspond to the singularities of eq. (\ref{kappa}) for $D=2$ and 3, respectively. The one-dimensional case is singular by definition.}
\item In the limit $q \to \infty$ ($\theta \to 2$) the entropy vanishes. This limit was used for constructing in a \emph{formal} way inflationary solutions with either an ordinary scalar field or a Dirac-Born-Infeld tachyon. However, perturbations of this background are frozen and the dynamics is trivial \cite{PhD}; this is consistent with $S_q=0$. Note that a redefinition of the entropy via a positive constant shift, $S_q\to S_q+S_0$ can easily solve the problem of negative states. But for the above argument, in the limit $\theta\to 2$ the action becomes singular and there is no univocal way to fix a nonvanishing contribution $S_0$. Therefore we can set $S_0=0$ consistently in the patch approximation.
\end{itemize}
Another application of eq. (\ref{S}) is related to the entropy bound proposed by Fischler and Susskind \cite{FS}, according to which the entropy of the universe does not exceed the dS entropy. In a flat universe, this corresponds to
\be
\sigma (aH)^{1-D} < S_q \sim H^{2-D-\theta},
\ee
where $\sigma={\rm const}$ is the comoving entropy density and $(aH)^{-1}$ is the comoving Hubble radius. For a power-law expansion, $a\sim t^n$, the horizon size is $H^{-1}\sim t$ and the above bound is always respected, provided
\be
n>\frac{1-\theta}{D-1}.
\ee
In four-dimensional general relativity one recovers the result of \cite{FS}, $n>1/3$. In the RS case this bound is trivial for an expanding universe, while in GB $n>2/3$. For a perfect fluid with equation of state $p=w\rho$, $1/n=\epsilon=(D-1)(1+w)/(2-\theta)$ and the bound becomes
\be
w<\frac{1}{1-\theta}.
\ee
While in general relativity this is consistent with the stiff matter bound, in the GB scenario the marginal case is $w=1/2$. See \cite{KL} for a discussion on entropy bounds in cosmology.

One can now show that the Friedmann equation is equivalent to the dS first law of event horizon thermodynamics, eq. (\ref{QTS}). The Gibbons-Hawking temperature of the de Sitter horizon does not depend on the gravitational action and reads \cite{GH1}
\be\label{T}
T=\frac{H}{2\pi}.
\ee
High-energy corrections like those we are considering are expected to play an important role in the inflation era, during which small deviations from the pure de Sitter background are not only allowed, but also necessary for explaining current large-scale observations. Therefore the above temperature (proportional to the surface gravity) is not constant but slowly decreasing in time.

The perfect fluid coupled to the gravity action (\ref{act}) has energy momentum $T^{\mu\nu} = \rho\,u^\mu\,u^\nu - p\,h^{\mu \nu}$, where $u^\mu$ is the $D$-velocity of the fluid and $h^{\mu \nu}=g^{\mu \nu}-u^\mu u^\nu$ is the spatial projector of the metric. During the quasi de Sitter regime typical of inflation such fluid can be constituted by a slow-rolling scalar field. The radial\footnote{In the cosmological context we can safely ignore an eventual angular momentum carried by matter and the flux integral associated to the Killing vector generating orthogonal rotations.} energy flux $\Phi$ through the horizon per unit area and time interval is then $\Phi=\rho+p$. In the models we are considering there is no appreciable brane-bulk exchange during inflation and the continuity equation for the perfect fluid is the same as in general relativity:
\be
\dot{\rho}+(D-1)H(\rho+p)=0.
\ee
Therefore the energy exchange through the area $A_{D-2}$ is
\be\label{E}
dE=-A_{D-2}\Phi dt=\frac{A_{D-2}}{(D-1)H}d\rho,
\ee
where the $-$ sign comes from the fact that one is measuring the incoming flux from the region outside the chosen surface (in our case the dS horizon) into the causally connected inner patch (inward pointing normal). In the black hole picture this flux describes an accretion process triggered by the infall of a small amount of matter from the asymptotically flat region into the black hole. 

Plugging all the above expressions into eq. (\ref{QTS}), one gets the Friedmann equation (\ref{FRW2}). It is straightforward to generalize this result to the case of eq. (\ref{Sf}).

All that has been said depends only weakly on the slow-roll approximation assumed from the beginning. In fact, when using eq. (\ref{Lg}) instead of the gravitational part of eq. (\ref{pact}) the coefficients in it conspire so that to give again the entropy (\ref{S}) at the desired slow-roll order, provided those coefficients are fixed as explained above. Corrections to the entropy will be very small during the inflationary regime, whereas near the end of inflation, when $\epsilon \sim 1$, the evolution of the effective action (\ref{pact}) and its thermodynamical description will break down. However, it may be possible that at that stage the standard general relativistic picture is already recovered.


\section{Discussion} \label{conc}

To summarize, we have shown that the effective Friedmann equation characterizing inflationary cosmologies which follow a nonstandard evolution (in modified gravity scenarios such as the braneworld) can be regarded as the first law of the thermodynamics of the de Sitter horizon of an effective gravitational theory. This picture can give important informations that are not available when considering only the dynamics of such systems. In particular, one can constrain the validity of the approximation $H^2\propto \rho^q$ and the range of the effective parameter $q$ in terms of properties of the effective entropy.

A topic which is still to be investigated is the behaviour of the cosmological perturbations generated by quantum fluctuations of the inflationary scalar field. From the perturbed first law of thermodynamics one can derive, along the lines of \cite{FrK}, the perturbation equation for the scalar field in the gravitational field described by eq. (\ref{pact}), and compare it with the linearized Einstein equations of the original theory. We expect that the result will be in agreement with past calculations (see \cite{PhD} for a treatment in the same framework and references therein for the full braneworld case).

Before concluding let us remark that the entropy (\ref{S}) associated to the high-energy modification of the Friedmann equation has a direct interpretation in terms of braneworld quantum tunneling. The birth probability $P(\phi_i)$ of the universe from ``nothing'' ($a=0$) to a state with $a=H^{-1}(\phi_i)$ for some initial matter configuration $\phi_i$ is described as the square of the Wheeler-DeWitt wave function \cite{tupr} 
\be
P(\phi_i)=e^{-2|S_{\rm E}|},
\ee
where $S_{\rm E}$ is the euclidean action associated to the tunneling process. In particular, this coincides with the de Sitter entropy of the system up to a $-$ sign, $S_{\rm E}=-S$. Also, the Hartle-Hawking probability $P_{\rm HH}(\phi)=\exp(-2S_{\rm E})$ \cite{HaH} describes the probability of realizing a quantum fluctuation to a configuration $\phi=\text{const}$ in a universe which has already been born.

One may ask whether the de Sitter entropy of the effective gravity (\ref{pact}) can give some insight into the creation of the brane universe or the inflationary self-reproduction process, respectively. For instance, in the first case $P\sim \exp[-C\rho^{(\theta+2)/(\theta-2)}]$, $C$ being a positive constant, which in the range given by eq. (\ref{range}) favours high-energy initial conditions. The high-energy Gauss-Bonnet regime is less favoured than both the high-energy Randall-Sundrum and 4D ones as initial state of the universe. Moreover, the most favoured scenario (high-energy RS) is the same for which there is no cosmological entropy bound for a power-law expansion. Very roughly, this might be the hint that indeed the best candidate in terms of initial conditions is general relativity (as a theory by itself or as a low-energy limit of the braneworld). All these would be notable results if it made sense to compare disjoint probabilities in different gravity models.

The birth of the braneworld was considered in the low-energy limit where the effect of the bulk back-reaction can be neglected \cite{GaS,KoS2}. The resulting tunnel probability is the same as in general relativity, the de Sitter entropy satisfying the usual area law.\footnote{The cited papers lie closer to the AdS/CFT spirit, where the ``low-energy'' regime corresponds to a brane with Hubble horizons larger than the AdS curvature scale $\ell H\ll 1$. In \cite{KoS2}, a brane with constant tension is in a Sc-AdS bulk hosting a black hole with small mass $M$. The area law is then spoiled explicitly when $M\neq 0$.} At high energies we should expect corrections to both the entropy and the probability function. We would like to propose the above model as the approximate description of such systems in the high-energy limit; a full computation in the original gravitational theories should check this conjecture explicitly.

As inflation damps away any contribution from the curvature and the projected Weyl tensor (the latter at least at large scales), we have neglected the effect of both. However, during the first instants of life of the universe these terms can be relevant, although for a short time. Also, at high energies the Weyl contribution becomes important at sub-horizon scales \cite{KMW}. In order to consistently connect de Sitter entropy and tunneling probability, it is important to clarify the drawbacks of the above assumptions.

As regards the curvature, one can extend the above procedure by using an apparent horizon $\tilde{H}^{-1}$ instead of the Hubble radius $H^{-1}$ in eqs. (\ref{S}), (\ref{T}), and (\ref{E}) \cite{BR}. The former is defined by the equation $h^{bc}\partial_b \tilde{r}_H\partial_c \tilde{r}_H=0$ in the metric (\ref{metric}), where the indices $b$ and $c$ denote the $t$ and $r$ coordinates and $h_{bc}={\rm diag}(-1,a^2/(1-kr^2))$. Therefore
\be\label{apphr}
\tilde{H}\equiv \tilde{r}_H^{-1}= \sqrt{H^2+\frac{k}{a^2}}.
\ee
We refer to \cite{BR} for a precise derivation of the first law of thermodynamics.

The influence of bulk physics on the effective four-dimensional evolution can be encoded in two modifications, the first of the Friedmann equation and the second one of the fluid continuity equation. In the latter case, there is an energy flux between the brane and the bulk so that the expansion of the universe is no longer adiabatic: $dE_{\rm tot}+pdV\neq 0$, where $E_{\rm tot}=\rho V$ is the total energy in the physical volume $V\sim a^3$. The continuity equation entered the derivation of the Friedmann equation from the first law of thermodynamics, so it is natural to expect substantial deviations, both conceptual and quantitative, from the above picture. 

If the bulk is empty there is no flux along the extra dimension, but still there can be some nonvanishing Weyl term in the evolution equation, encoded in an effective energy density $\rho_{\cal E}(t)$. In the Randall-Sundrum scenario, $\rho_{\cal E}\sim a^{-4}$.

We can deduce that the simplest holographic interpretation of the Friedmann equation in terms of thermodynamical processes might be typical of maximally symmetric spaces but not hold in more complicated backgrounds like those at the basis of the braneworld.\footnote{See also the parallel discussion of \cite{MyK} in the context of the Cardy-Verlinde formula.} One can define the effective curvature $\tilde{k}\equiv k-\rho_{\cal E}(t)a^2(t)$ and replace $k\to \tilde{k}$ in eqs. (\ref{metric}) and (\ref{apphr}). For a flat universe and a positive definite Weyl density, the apparent horizon, which is always well defined for a monotonically evolving universe, is larger than the Hubble radius. The resulting metric is no longer FRW and describes an ``onion''-like background which is isotropic but not homogeneous, so that the Einstein equations will depend on $r^2$. However, it is easy to check, via the thermodynamical approach and at least in the RS case, that one indeed gets the braneworld Friedmann equation with the correct Weyl contribution at the apparent horizon. Therefore there is a sort of correspondence between the full theory and the horizon thermodynamics of an effective theory in an effective background. But the original beauty of the holographic principle, relating different aspects of one theory in a given background, seems completely lost. Further studies of the high-energy braneworld, more complete than that presented here, are of course required.


\acknowledgments
I thank K. Koyama, K. Maeda, N. Okuyama, A. Padilla, M. Sasaki, T. Tanaka, and T. Torii for useful discussions. This work is supported by JSPS.


\end{document}